\documentclass[prl,twocolumn,superscriptaddress,longbibliography,floatfix]{revtex4-2}
\usepackage{graphicx,amsfonts,amssymb,amsmath,xspace,dsfont}
\usepackage[colorlinks=true,citecolor=BurntOrange,linkcolor=red,urlcolor=Bittersweet]{hyperref}
\usepackage{color}
\usepackage[dvipsnames]{xcolor}
\definecolor{g}{rgb}{.1,0.4,.1} 
\definecolor{b}{rgb}{0,0.2,1}
\definecolor{rouge}{rgb}{0.82,0.,0.}
\definecolor{vert}{rgb}{0.,0.82,0.}
\definecolor{orange}{rgb}{1,0.5,0.}
\definecolor{bleu}{rgb}{0.,0.,0.82}
\definecolor{m}{rgb}{0.82,0.,0.82}
\definecolor{vert2}{rgb}{0.,0.5,0.}
\definecolor{rougeclair}{rgb}{1.0,0.7,0.7}

\newcommand{\be}{\begin{equation}}
\newcommand{\ee}{\end{equation}}
\newcommand{\beqn}{\begin{eqnarray}}
\newcommand{\eeqn}{\end{eqnarray}}

\newcommand{\tr}{\mathrm{Tr}}

\begin{document}

\title{Kitaev model on Hurwitz hyperbolic tilings: A non-Abelian gapped chiral spin liquid}

\author{R\'emy Mosseri}
\email{remy.mosseri@sorbonne-universite.fr}
\affiliation{Sorbonne Universit\'e, CNRS, Laboratoire de Physique Th\'eorique de la Mati\`ere Condens\'ee, LPTMC, F-75005 Paris, France}

\author{Yasir Iqbal}
\email{yiqbal@physics.iitm.ac.in}
\affiliation{Department of Physics and Quantum Centre for Diamond and Emergent Materials (QuCenDiEM), Indian Institute of Technology Madras, Chennai 600036, India}

\author{Roger Vogeler}
\email{vogelerrov@ccsu.edu}
\affiliation{Department of Mathematical Sciences, Central Connecticut State University, New Britain, Connecticut 06050, USA}

\author{Julien Vidal}
\email{julien.vidal@sorbonne-universite.fr}
\affiliation{Sorbonne Universit\'e, CNRS, Laboratoire de Physique Th\'eorique de la Mati\`ere Condens\'ee, LPTMC, F-75005 Paris, France}
\affiliation{Department of Physics and Quantum Centre for Diamond and Emergent Materials (QuCenDiEM), Indian Institute of Technology Madras, Chennai 600036, India}

\begin{abstract}
We study the Kitaev model on the trivalent heptagonal Hurwitz hyperbolic tiling. The presence of odd-length loops in the tiling is responsible for a spontaneous time-reversal symmetry breaking. Interestingly, at the isotropic point, the two degenerate ground states (Kramers pair) are shown to be gapped chiral spin liquids and the elementary excitations are non-Abelian Ising anyons.
\end{abstract}

\maketitle

{\it Introduction.} In 1987, Kalmeyer and Laughlin established an equivalence between resonating-valence-bond and fractional quantum Hall states~\cite{Kalmeyer-1987}. This important achievement gave rise to a new paradigm: (Abelian) chiral spin liquids (CSLs) --- quantum spin liquids~\cite{Zhou-2017} that break reflection symmetry and time-reversal symmetry (TRS). The lure lay in hosting deconfined spinon and holon excitations, and thus being a stable zero-temperature phase. By giving birth to the notion of topological quantum order~\cite{Wen-1990,Wen-1991}, CSLs played a major role in higher echelons of condensed matter physics. It was not long after that other fractional quantum Hall states \cite{Moore-1991,Wen-1991b} further unveiled the possibility of chiral topological phases featuring exotic excitations with non-Abelian quantum statistics~\cite{Goldin-1985}.

In general, finding parent Hamiltonians for quantum spin liquids is a nontrivial task, and proves to be especially intricate for non-Abelian CSLs~\cite{Greiter-2014}. In this endeavour, the Kitaev model~\cite{Kitaev06}  appeared as a rare example of a system exhibiting both chiral and achiral topological phases while being exactly solvable. Although it may not be sufficient, the simplest mechanism to obtain chiral phases is to break TRS. In the Kitaev model, this can be achieved either explicitly (e.g., by introducing a magnetic field) or spontaneously (if the two-dimensional system considered contains some odd cycles. Since the Kitaev model is defined on a trivalent graph, the latter scenario cannot be realized in the Euclidean plane for regular tilings (the only regular trivalent graph in flat space being the honeycomb lattice).
Hence, to study the role of spontaneous TRS breaking in this model, one should either consider nonregular tilings (see, e.g., Refs.~\cite{Yao07,Dusuel08_2,Peri-2020,Cassella23,Grushin23}) or regular tessellations of two-dimensional surface with constant negative curvature~\cite{Dusel24}, i.e., hyperbolic tilings (see Ref.~\cite{Boettcher22} for a recent review). 

Recent explorations of quantum matter in hyperbolic geometry have been devoted to developing a hyperbolic band theory for non-interacting systems~\cite{Maciejko-2021,Cheng-2022,Maciejko-2022,Lenggenhager23}, and their understanding within the purview of symmetry protected topological phases~\cite{Stegmaier22,Tummuru-2024,Urwyler-2022}. Of late, the focus has elevated toward understanding quantum many-body phenomena through metal-insulator transitions and magnetism~\cite{gotz-2024}, fractional Chern insulators~\cite{he-2024}, and lattice-regularized formulations of AdS-CFT correspondence~\cite{Basteiro-2023,Dey-2024}. Interestingly, quantum and classical circuit networks have lent themselves as accessible platforms for experimental realization of hyperbolic space~\cite{Kollar-2019,Lenggenhager-2022,Chen-2023}.

%
%
\begin{figure}[t]
\centering
\includegraphics[width=.92\columnwidth]{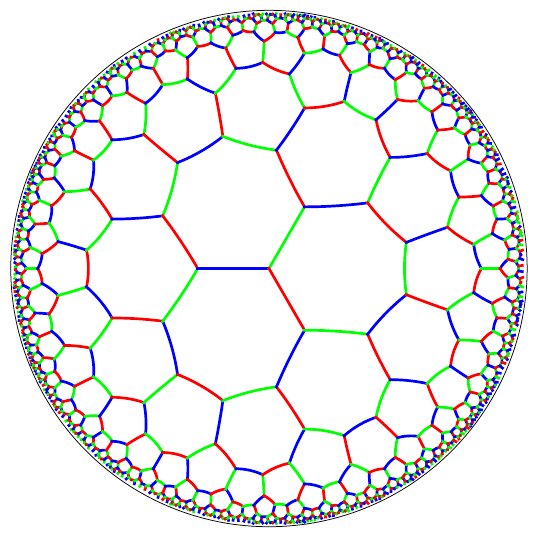}
\caption{A possible three-edge coloring of the $\{7,3\}$ tiling with $N_{\rm v}=1162$ vertices in the Poincar\'e disk conformal representation.}
\label{fig:disk}
\end{figure}
%
%

In this Letter, we study the Kitaev model on the $\{7,3\}$ hyperbolic tiling which is a trivalent regular graph made of heptagonal plaquettes (see Fig.~\ref{fig:disk}). Unlike recent studies on other regular hyperbolic tilings~\cite{Lenggenhager24,Dusel24}, the two degenerate ground states (Kramers pair) are gapped CSLs which spontaneously break TRS at the isotropic point. To characterize this gapped phase, we compute the Chern number and find that it corresponds to an Ising (non-Abelian) chiral topological  phase. We also argue that the vortex (vison) excitations are the low-energy excitations of the system. Interestingly, we also find that the spectral weights are different in each band of the density of states (DOS), which is in stark contrast with the Euclidean case. 

{\it Model and symmetries.} For any three-edge colorable tiling embedded on an orientable surface, the Kitaev model Hamiltonian is given by
%
%
\begin{equation}
  \label{eq:ham0}
  H=-\sum_{(j,k)}
  J_{\alpha} \, \sigma_j^\alpha\sigma_k^\alpha,
\end{equation} 
%
%
where $\sigma_i^\alpha$ are the usual Pauli matrices acting on spin located on site $i$. In Eq.~\eqref{eq:ham0}, the sum runs over all link $(j,k)$ of the tiling and the type $\alpha$ is given by the corresponding edge color. As shown by Kitaev~\cite{Kitaev06}, for any plaquette $p$, the operator 
%
%
\begin{equation}
W_p= \prod_{(j,k)\in {\partial p}} \sigma_j^\alpha\sigma_k^\alpha,
\label{eq:wpdef}
\end{equation} 
%
%
is a conserved quantity, i.e., $[H,W_p]=0$. Here, the product runs over all links $(j,k)$ of the oriented plaquette $p$ (e.g., clockwise). Moreover, one has  $[W_p,W_p']=0$, so that one can block diagonalize the Hamiltonian according to each map of $W_p$'s. 

For a given configuration of the $W_p$'s, replacing Pauli matrices by Majorana operators, the Hamiltonian \eqref{eq:ham0} can be written as
%
%
\be 
H=\frac{\mathrm{i}}{4} \sum_{j,k} A_{jk} c_j c_k,
\label{eq:ham_Majo}
\ee  
%
%
where the sum runs over all sites of the tiling, and where  $c_j$ is a Majorana operator acting on site $j$. As explained by Kitaev, this mapping presumes an extended Hilbert space so that a projection onto the physical subspace must be ultimately performed~\cite{Kitaev06}.  
The matrix $A$ is a real skew-symmetric matrix whose elements depend on $\mathbb{Z}_2$ variables \mbox{$u_{jk}=-u_{kj}=\pm 1$}, defined on each link of the lattice. More precisely, if $j$ and $k$ are connected by a link of type $\alpha$,  $A_{jk}=2 \, J_{\alpha} \, u_{jk}$ (see Ref.~\cite{Kitaev06} for a detailed derivation). In this language, the eigenvalue of the operator $W_p$ is given by
%
%
\be
w_p=(-{\rm i})^q \prod_{(j,k)\in {\partial p}} u_{jk},
\label{eq:fluxdef}
\ee
%
%
where $q$ is the number of edges of the plaquette $p$, and where, again, the product runs over all links $(j,k)$ of the oriented plaquette $p$~\cite{Kitaev06,Petrova14}.
Hence, for each configuration of the gauge variables $u_{jk}$'s, the problem is reduced to a quadratic Majorana fermion Hamiltonian. Let us stress that, on a closed surface, one always has $\prod_p w_p=1$, or equivalently $\prod_p W_p=\mathds{1}$, which reflects the well-known total flux quantization constraint for a closed oriented surface~\cite{Dirac31}.

Importantly, the Hamiltonian \eqref{eq:ham0} is time-reversal symmetric. However, if the tiling contains plaquettes of odd length (odd $q$), TRS is spontaneously broken~\cite{Kitaev06}. This phenomenon has been first analyzed in the decorated honeycomb lattice~\cite{Yao07,Dusuel08_2}. \\

{\it Three-edge coloring.} In this work, we consider the Kitaev model~\cite{Kitaev06} on the regular $\{7,3\}$ hyperbolic tiling (see Fig.~\ref{fig:disk}). This trivalent tiling (planar cubic graph) is one of the regular  tilings that can tile Hurwitz surfaces~\cite{Hurwitz92} (see also Refs.~\cite{Conder90,Vogeler_Thesis} for more recent studies). As a bridgeless planar cubic graph, this tiling can be three-edge colored (see Fig.~\ref{fig:disk} for a concrete example). Such a coloring, also known as Tait coloring~\cite{Tait80_1,Tait80_2,Tutte66,Robertson97}, ensures that one can still map the standard Kitaev spin-1/2 model onto a quadratic Majorana fermion model in a $\mathbb{Z}_2$ gauge field, as described above~\cite{Kitaev06}. 

For a given tiling, the phase diagram of the Kitaev model depends on the coloring considered (see, e.g., Ref.~\cite{Kamfor10}). In the absence of canonical coloring of the $\{7,3\}$ tiling, we focus, in the following, on the so-called isotropic point ($J_{\alpha}=J$ for all links) for which all three-edge colorings are equivalent. Without loss of generality, we  set  $J=1$. \\

{\it Ground-state sector(s).} Among all possible gauge choices, we look for the one which gives the lowest energy~\cite{Kitaev06}, 
%
%
\be
E_0=-\frac{1}{4} \tr |{\rm i}A|.
\label{eq:gsedef}
\ee
%
%
The corresponding ground-state manifold has been recently conjectured to be obtained when
%
%
\be
w_p^{{\rm g.s.}}=-(\pm {\rm i})^q,
\label{eq:fluxmin}
\ee
%
%
for each elementary plaquette $p$ with $q$ edges, provided the system is not too small~\cite{Cassella23}. This conjecture early proposed by Lieb and Loss~\cite{Lieb93}, and recently built up from sampling of amorphous systems (see also Ref.~\cite{Grushin23}), is notably in agreement with existing results on the regular $\{6,3\}$~\cite{Kitaev06,Lieb94}, $\{8,3\}$~\cite{Lenggenhager24}, and $\{9,3\}$~\cite{Dusel24} tilings. Interestingly, for $\{{\rm odd} \: p,3\}$ tilings, it predicts a two-fold degenerate ground-state manifold (i.e., a Kramers pair~\cite{Kramers30} associated with $w_p^{{\rm g.s.}}=\pm {\rm i}$), which is reminiscent of the spontaneous TRS breaking expected in this case. For such homogeneous configurations of $w_p$'s, the spectrum of the Majorana fermion Hamiltonian \eqref{eq:ham_Majo} is the same as the one computed in Ref.~\cite{Stegmaier22} for the Hofstadter butterfly with a flux per plaquette $\Phi$ (measured in units of the flux quantum) given $w_p={\rm e}^{{\rm i} \Phi}$. 
As a side remark, let us stress that, in the isolated-dimer limit, where one of the couplings dominates (e.g., $J_z \gg J_x,J_y$), the conjecture \eqref{eq:fluxmin} is corroborated by the perturbative result derived in Ref.~\cite{Petrova14}. \\
%
%
\begin{figure}[t]
\centering
\includegraphics[width=\columnwidth]{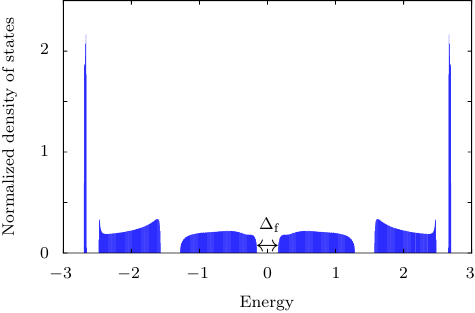}
\caption{Normalized DOS of the matrix ${\rm i}A/2$ in the ground-state sectors \mbox{$w_p=\pm{\rm i}$} obtained from ED of our optimal PBC cluster with \mbox{$N_{\rm v}=152 096$} vertices (bin width=0.005). From left to right, the six energy bands  have the following spectral weights: $1/14, 3/14, 3/14, 3/14,3/14, 1/14$. A finite fermionic gap $\Delta_{\rm f} \simeq 0.31$ is observed around $E=0$.}
\label{fig:DOS}
\end{figure}
%
%

{\it Method.} Despite some recent developments of an hyperbolic band theory~\cite{Maciejko-2021,Boettcher22}, the noncommutativity of translations in the hyperbolic plane prevents a full analytical calculation of the Hamiltonian \eqref{eq:ham_Majo}. Consequently, one must use numerical tools to compute the energy spectrum, and one then has the choice between several strategies. The first one consists in computing the DOS using either the continued-fraction~\cite{Mosseri23} or the supercell~\cite{Lenggenhager23} method. Alternatively, one may perform exact diagonalizations (ED) of the Hamiltonian defined on clusters with periodic boundary conditions (PBC) to spurious boundary effects. As far as the spectrum is concerned, these approaches are complementary~\cite{Stegmaier22,Lenggenhager23,Lenggenhager24,Dusel24}. In this work, we focus on clusters with PBC (regular tessellations of compact Hurwitz surfaces) which allows one to compute very efficiently the Chern number $\nu$ using the Widom-St$\check{\rm{r}}$eda formula~\cite{Streda82,Widom82} (see Ref.~\cite{Mosseri22} for a related study) by adding an extra uniform magnetic field. In the present context, the Chern number determines the nature of the topological phase according to the sixteenfold way classification~\cite{Kitaev06}. 

Contrary to the Euclidean case, the construction of periodic hyperbolic regular tilings is a nontrivial problem since the genus $g$ of the surface increases with the number of vertices $N_{\rm v}$. More precisely, for $\{p,3\}$ tilings, one has $N_{\rm v}= 4\, p \, (g-1)/(p-6)$. For the special case $p=7$, these periodic tilings are associated to Hurwitz surfaces~\cite{Hurwitz92} (hence their name), the genus of which follows an infinite sequence~\cite{HurwitzOEIS}.
We construct the tilings computationally using GAP~\cite{GAP4,GRAPE}. The symmetry group of a tiling (i.e., a Hurwitz group) is first specified as a quotient of the (2,3,7) triangle group by means of a group presentation. An isomorphic permutation group is then produced, and its Cayley graph is generated.  Modification of the Cayley graph then yields the actual 1-skeleton of the $\{7,3\}$ tiling as a list of adjacencies.

Since one can only diagonalize finite-size clusters, one has to check the convergence of the results with respect to several  parameters. Indeed, for a given cluster, the spectrum of $H$ depends on (i) its size, (ii) its shape  [gluing conditions (PBC) may be different for a fixed system size], and (iii) the gauge choice which inserts Aharonov-Bohm (AB) fluxes inside the $2g$ noncontractible closed loops of the tiling (see Refs.~\cite{Stegmaier22,Mosseri22} for related discussions). By comparing with the moments of $H$ obtained on large clusters with open boundary conditions)~\cite{Mosseri23}, we found that the first 34 moments of $H$ are exactly reproduced by the largest PBC clusters considered here  containing $N_{\rm v}=152 096$ (the $n$-th moment of $H$ is defined as $\frac{1}{N_{\rm v}} {\rm Tr}  \: H^n$).\\

{\it Results.} The DOS of the $\{7,3\}$ tiling in the ground-state sector is displayed in Fig.~\ref{fig:DOS}. This symmetric DOS is made of six bands separated by gaps. We computed the weights of these bands and found that, contrary to the Euclidean case,  these depend on the bands (see caption of Fig.~\ref{fig:DOS}). We checked that these results are stable when changing the system size. This is a genuine feature of the hyperbolic tilings which, again, shows, that the recently proposed hyperbolic band theory based on one-dimensional irreducible representation of the Fuchsian (translation) group is widely incomplete~\cite{Maciejko-2021}. A deeper understanding of these spectral weights is beyond the scope of the present work but it definitely deserves further attention.

From the energy spectrum, one can easily compute the ground-state energy \eqref{eq:gsedef} per vertex. As can be seen in Table~\ref{tab:results}, this energy lies between the one of the $\{6,3\}$ tiling and those of the $\{\infty,3\}$. However, the most salient feature of this spectrum is the presence of a finite fermionic gap $\Delta_{\rm f} \simeq 0.31$ at $E=0$ (see Fig.~\ref{fig:DOS}). Although a gap is generically found for amorphous systems~\cite{Grushin23,Cassella23} breaking spontaneously TRS, this is the unique known example of a gapped spin liquid observed in the Kitaev model for a regular tiling at the isotropic point. Indeed, all other $\{p,3\}$ tilings studied so far ($p=6$~\cite{Kitaev06}, $8$~\cite{Lenggenhager24},  and $9$~\cite{Dusel24}), are gapless in the ground-state sector at the isotropic point (see Table~\ref{tab:results}). We also checked that it is the case for larger $p$ as well as for $p=\infty$~\cite{Mosseri22}. From that respect, the case $p=7$ must be considered as singular. In addition, since TRS is spontaneously broken for this tiling, one expects a chiral topological phase in the neighborhood of this point. We computed the Chern number using the method described in Ref.~\cite{Mosseri22} and we found that $\nu=\pm 1$ [the sign depends on the choice of the ground-state sector ($w_p^{{\rm g.s.}}=\pm {\rm i}$]. 
As a result, one gets a gapped CSL associated to an Ising topological phase (non-Abelian anyon theory)~\cite{Kitaev06}. 
\begin{table}[h]
\center
\begin{tabular}{| l | l | l | l | }
\hline
Tiling & $E_0/N_{\rm v}$ & $\Delta_{\rm f}$ & $\Delta_{\rm v}$  \\
\hline
\hline  
$\{ 6,3 \}$ & -0.787299~\cite{Kitaev06} & 0 & 0.26~\cite{Kitaev06} \\
\hline  
$\{ 7,3 \}$ & -0.7684 &  0.31 & \hspace{-1mm} 0.13  \\
\hline  
$\{ 8,3 \}$ & -0.7692~\cite{Lenggenhager24}& 0 & 0.08  \\
\hline  
$\{ 9,3 \}$ & -0.7638~\cite{Dusel24} & 0 & 0.03~\cite{Dusel24}  \\
\hline  
$\{ \infty,3 \}$ & -0.762734 & 0 & 0  \\
\hline
\end{tabular}
\caption{Summary of known results for the Kitaev model on regular trivalent tilings at the isotropic point ($J=1$). For the $\{ 6,3 \}$ and $\{ \infty,3 \}$ tilings, $E_0/N_{\rm v}$ can be computed analytically in the thermodynamical limit as an integral, using the exact DOS. Similarly, an analytical expression of the vison gap for the $\{ 6,3 \}$ tiling has been recently derived in Ref.~\cite{Panigrahi23}.}
\label{tab:results}
\end{table}
%
%

For a topological phase described by a unitary modular tensor category $\mathcal{C}$ defined on a closed orientable manifold of genus-$g$, the ground-state degeneracy is given by the Moore-Seiberg-Banks formula~\cite{Moore89}:
%
%
\be
D_0=\sum_{A \in \mathcal{C}} \left(\frac{d_A}{\mathcal{D}}\right)^{2-2g},
\ee
%
%
where $d_A$ is the quantum dimension of the object $A$ and \mbox{$\mathcal{D}=\sqrt{\sum_A d_A^2}$} is the total quantum dimension. For the Ising category, one has three objects usually denoted $1,\sigma$, and $\psi$ with quantum dimension $d_1=1$, $d_\sigma=\sqrt{2}$, and $d_\psi=1$ (see,  e.g., Refs.~\cite{Kitaev06,Rowell09}), so that the total ground-state degeneracy is given by
%
%
\be
D_0=\frac{1}{2}(2^g+4^{g}).
\ee
%
%
The reduction from the total number of different AB flux sectors, $2^{2g}$, to $D_0$ stems from the projection procedure onto the physical Hilbert space. However, for the problem at hand and because TRS is spontaneously broken, one must add an extra factor of two coming from the Kramers degeneracy so that the ground-state degeneracy is  $2 D_0$ [a related discussion can be found in Ref.~\cite{Yao07} for the decorated honeycomb lattice on a torus ($g=1$)]. 
We emphasize that for a finite-size system, which, for hyperbolic tilings, means a finite $g$, this degeneracy might be partially lifted by the AB fluxes. This splitting is expected to vanish as ${\rm e}^{-\# L}$, where $L$ is the length of the shortest noncontractible loop (systole). In hyperbolic tilings, one typically has  $L\propto \log N_{\rm v}$. Hence, one expects the splitting to vanish algebraically with the system size.  

Finally, we computed the vison gap obtained by flipping one link ($u_{jk}\rightarrow -u_{jk}$). Such a flip excites two adjacent plaquettes changing the sign of the corresponding $w_p$'s [see Eq.~\eqref{eq:fluxdef}]. These two excited plaquettes may be further separated by flipping an {\it ad hoc} sequence of $u_{jk}$. We checked the emergence of two zero-energy modes when one adds two well-separated visons in the system, as expected for odd Chern numbers~\cite{Kitaev06}.
As recently observed in the $\{9,3\}$ tiling, we found that the two-vison lowest-energy configuration corresponds to adjacent excited plaquettes for which the vison gap is $\Delta_{\rm v}=E_{2 \rm v}-E_0\simeq 0.13$ (see Table~\ref{tab:results} for a comparison with other regular trivalent tilings). The main contribution to this gap comes from twelve bound states localized around the two vortices. These bound states are easily identified by computing, e.g., the inverse participation ratio of the corresponding eigenstates. Interestingly, in our case, the vison gap is lower than the fermion gap ($\Delta_{\rm v}<\Delta_{\rm f}$). As a consequence, low-energy excitations of $H$ are (very likely) adjacent pairs of visons although we cannot firmly rule out the possibility that other vortex configurations have a lower energy cost.\\

{\it Summary and outlook.} We analyze the Kitaev model at the isotropic (coloring-independent) point in the $\{7,3\}$ tiling by means of ED of large PBC clusters. The two-fold degenerate ground states (Kramers pairs) are found to be gapped CSLs  characterized by a Chern number $\nu=\pm1$. This surprising result is in stark contrast with all other regular trivalent tilings studied so far which are gapless at the isotropic point. In addition, we found that the vison gap is lower than the fermion gap suggesting that low-energy excitations are (adjacent pairs of) visons. Additional information about the role of a three-spin term~\cite{Kitaev06} explicitly breaking  TRS can be found in Ref.~\cite{Supp_Mat}.

An interesting perspective would be to analyze this model at finite temperature $T$. Indeed, although topological order is known to be robust against perturbations at $T=0$ in two dimensions, it is also known to be fragile in deconfined phases. For instance, Hastings showed that, for local commuting projector Hamiltonians, topological order is destroyed for any $T>0$, in the thermodynamical limit~\cite{Hastings11}. 
Similarly, in the honeycomb lattice, no finite-$T$ phase transition is observed~\cite{Nasu15,Eschmann_thesis}. Nevertheless, it has been shown that a finite-$T$ transition associated to TRS breakdown could exist in the Kitaev model when the system spontaneously breaks TRS at $T=0$~\cite{Nasu15}, as is the case for the $\{{\rm odd} \: p,3\}$ tilings. Thus, a detailed study of the finite-$T$ properties of the Kitaev model in $\{p,3\}$ tilings both in two and higher dimensions~\cite{Brien-2016,Eschmann-2020} would be of highly valuable. \\

\acknowledgments
 We thank G. Baskaran, F. Dusel, J.-N. Fuchs, T. Hofmann, P. Lenggenhager, A. Maity, and R. Thomale for fruitful discussions. The work of Y.I.\ was performed, in part, at the Aspen Center for Physics, which is supported by National Science Foundation Grant No.~PHY-2210452. The participation of Y.I. at the Aspen Center for Physics was supported by the Simons Foundation (1161654, Troyer). This research was supported in part by grant NSF PHY-2309135 to the Kavli Institute for Theoretical Physics (KITP) and the International Centre for Theoretical Sciences (ICTS), Bengaluru through participating in the program - Kagome off-scale (code: ICTS/KAGOFF2024/08). Y.I.\ acknowledges support from the ICTP through the Associates Programme, from the Simons Foundation through grant No.~284558FY19, and IIT Madras through the Institute of Eminence (IoE) program for establishing QuCenDiEM (Project No.~SP22231244CPETWOQCDHOC). Y.I. and J.V. acknowledge the use of the computing resources at HPCE, IIT Madras. J.V. thanks IIT Madras for a Visiting Faculty Fellow position under the IoE program during which this work was initiated and completed.


%

\appendix
\onecolumngrid
\vspace{12mm}

\begin{center}
\textbf{SUPPLEMENTAL MATERIAL}
\end{center}

\section{Role of a three-spin interaction term}
\label{appx:flux_sector_investigation}

In his seminal paper, Kitaev~\cite{Kitaev06} has shown that adding a three-spin term in the Hamiltonian~\eqref{eq:ham0} explicitly breaks TRS while keeping a quadratic form for the Majorana Hamiltonian~\eqref{eq:ham_Majo}. More precisely, this three-spin term proportional to $\kappa$ generates next-nearest-neighbor hoppings, $A_{jk}=2 \, \kappa \, u_{jl} \, u_{lk}$, where $l$ is the site connecting $j$ and $k$. By convention, we choose here  to orientate clockwise the triplet $(k,l,j)$~\cite{Kitaev06}.

At $\kappa=0$, as explained in the main text, the presence of odd cycles (heptagonal plaquettes) in the $\{7,3\}$ tiling spontaneously breaks  TRS and gives rise to a Kramers pair. However, since $\kappa$ breaks explicitly TRS, it lifts the degeneracy between the two ground states obtained for either $w_p=+{\rm i}$ or $w_p=-{\rm i}$. Here, our goal is to discuss the influence of $\kappa$ on both the fermionic and the vison gaps and to study this degeneracy lifts. 

\begin{figure}[h]
\centering
\includegraphics[width=0.49\columnwidth]{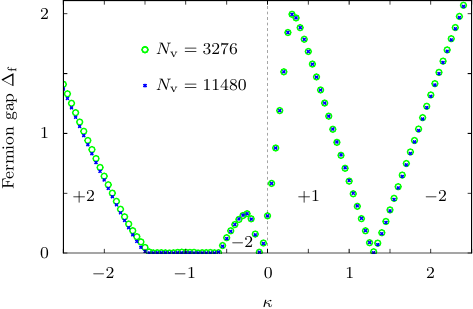} \hfill
\includegraphics[width=0.49\columnwidth]{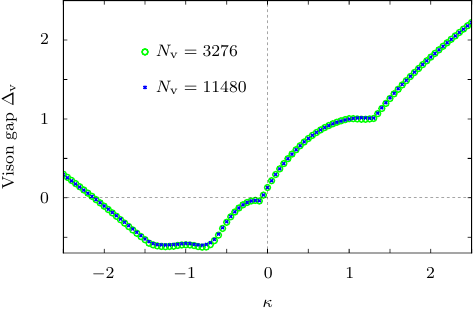}
\caption{Fermion (left) and vison (right) gaps  as a function of $\kappa$ for two different PBC clusters in the sector where $w_p=+{\rm i}, \forall p$. The Chern number $\nu$ is indicated in each gapped phases. Results for the other sector ($w_p=-{\rm i}, \forall p$) are obtained by a mirror symmetry with respect to $\kappa=0$ and by changing the sign of Chern numbers. Dashed lines are guides for the eyes. }
\label{fig:kappavisongap}
\end{figure}

Let us start by analyzing the role of $\kappa$ in one of these two ground-state sectors, e.g., $w_p={\rm i}$. As can be  seen in Fig.~\ref{fig:kappavisongap} (left), the fermionic gap $\Delta_{\rm f}$ displays a rich behavior when $\kappa$ is varied.
%
%

\noindent Indeed, we found four gapped phases characterized by different Chern numbers, separated by either gapless points located at $\kappa\simeq -0.1,+1.3$, or by a gapless phase found in the range $\kappa \in [-1.5,-0.6]$.
The gapped phases are characterized by different Chern numbers and associated to different topological phases according to the sixteen-fold way classification~\cite{Kitaev06}: $\nu=\pm 1$ corresponds to an Ising (non-Abelian) phase, whereas $\nu=\pm 2$ corresponds to a $\mathbb{Z}_4$ (Abelian) phase. Note that the existence of such a gapless phase in the Kitaev model has been first reported in a decorated square lattice~\cite{Baskaran09} and more recently in the hyperbolic $\{9,3\}$ lattice~\cite{Dusel24} (see also Ref.~\cite{Fuchs20} for similar results in the honeycomb lattice with triangular vortex configurations). 

We also computed the vison gap for two adjacent plaquettes as a function of $\kappa$ [see Fig.~\ref{fig:kappavisongap} (right)]. As can be seen, $\Delta_{\rm v}$ becomes negative for $\kappa\in[-2.1,-0.1]$ suggesting that the ground-state sector could be destabilized in this range. 

To go beyond, one must analyze the competition between the two ground-state sectors corresponding to $w_p=\pm {\rm i}$. As explained above, $\kappa$ lifts this degeneracy. So, for a fixed value of $\kappa$, one must determine which of these two sectors has the lowest energy. We computed the energy difference between the two sectors as a function of $\kappa$ and we found that for $\kappa>0$ ($\kappa<0$), the sector $w_p= {+ {\rm i}}$ ($w_p= {- {\rm i}}$) has always a lower ground-state energy. 
As a result, the actual fermion and the vison gaps are symmetric with respect to $\kappa=0$ (as could have been anticipated). For $\kappa>0$, they are given by the value shown in Fig.~\ref{fig:kappavisongap}.

However, one cannot exclude that other vortex sectors, i.e., other maps of $w_p$'s, have lower energies. We simply point out that the robustness of the topological phase found at $\kappa=0$ ($\nu=\pm 1$, ensures its stability in the neighborhood of this point. 

\end{document}